\documentclass[submission,copyright,creativecommons]{eptcs}
\providecommand{\event}{Non-Classical Logics 2024 (NCL'24)}

\usepackage{proof}

\usepackage{color}

\newcommand{\PROOF}{\mbox{\bf Proof.}\ \ }
\newcommand{\QED}{\hfill \rule{1.3ex}{0.6em}}
\newcommand{\SEQ}[2]{\mbox{$ #1 \Rightarrow #2 $}}
\newcommand{\I}{\mbox{$\to$}} 
\newcommand{\LAND}{\mbox{$\land$}} 
\newcommand{\LOR}{\mbox{$\lor$}}

\newcommand{\al}{\mbox{\it $\alpha$}}
\newcommand{\be}{\mbox{\it $\beta$}}
\newcommand{\ga}{\mbox{\it $\gamma$}}
\newcommand{\de}{\mbox{\it $\delta$}}

\newcommand{\GA}{\mbox{\it $\Gamma$}}
\newcommand{\DE}{\mbox{\it $\Delta$}}
\newcommand{\SI}{\mbox{\it $\Sigma$}}
\newcommand{\PI}{\mbox{\it $\Pi$}}

\newcommand{\M}{\mbox{$\sharp$}} 
\newcommand{\G}{\mbox{\rm G}}
\newcommand{\F}{\mbox{\rm F}}
\newcommand{\X}{\mbox{\rm X}}

\newtheorem{prop}{Proposition}[section] 
\newtheorem{thm}[prop]{Theorem} 
\newtheorem{lm}[prop]{Lemma} 
 
\newtheorem{df}[prop]{Definition} 
 
\newtheorem{rmk}[prop]{Remark}

\usepackage{iftex}

\ifpdf
  \usepackage{underscore}         
  \usepackage[T1]{fontenc}        
\else
  \usepackage{breakurl}           
\fi

\title{A Unified Gentzen-style Framework for Until-free LTL}
\author{Norihiro Kamide
\institute{Nagoya City University, Japan}
\email{drnkamide08@kpd.biglobe.ne.jp}
\and
Sara Negri
\institute{University of Genoa, Italy}
\email{sara.negri@unige.it}
}
\def\titlerunning{A Unified Gentzen-style Framework for Until-free LTL}
\def\authorrunning{Norihiro Kamide and Sara Negri}
\begin{document}
\maketitle

\begin{abstract}
A unified Gentzen-style framework for until-free propositional linear-time temporal logic is introduced. The proposed framework, based on infinitary rules and rules for primitive negation,  can handle uniformly both a single-succedent sequent calculus and a  natural deduction system.  Furthermore, an equivalence between these systems, alongside with proofs of cut-elimination and normalization theorems, is established. 
\end{abstract}

\section{Introduction}

Linear-time temporal logic (LTL) and its fragments and variants 
have been studied extensively 
\cite{PNUELI1977,KAWAI87,EMERSON1990,BM2003,BM04,BBGS2006,KAMIDEBSL06,GHLNO-CSL-2008,BL-2008,BN-2009,BN-2010,KW-FI-2011,KAMIDEMSCS2015,CGP-FSCD-2023}. 
In particular, many of Gentzen-style sequent calculi for LTL and its until-free fragment have been introduced and investigated  \cite{KAWAI87,PAECH1988,PLIUSKEVICIUS1991,SZABO1980,BM04,KAMIDEBSL06,GHLNO-CSL-2008,BL-2008,KAMIDEMSCS2015}. Some natural deduction systems for LTL and its until-free fragment have also been introduced and investigated \cite{BM2003,BBGS2006}. This study considers the until-free propositional fragment of LTL as a target logic. A reason for considering this fragment is that it is highly compatible with Gentzen's sequent calculus and natural deduction systems, LJ and NJ, \cite{GENTZEN,PRAWITZ} for intuitionistic logic. Namely, the proposed Gentzen-style sequent calculus and Gentzen-style natural deduction system for the fragment can be obtained as modified extensions of LJ and NJ, respectively.

Gentzen-style sequent calculi for LTL have been considered previously in the literature.
A sequent calculus LT$_{\omega}$ was introduced by Kawai for first-order until-free LTL, and  cut elimination and completeness  were proved \cite{KAWAI87}. 
 A 2-sequent calculus 2S$\omega$ for first-order until-free LTL,  with a cut elimination and a completeness proved were given by Baratella and Masini \cite{BM04}. An equivalence theorem between the propositional fragments of LT$_{\omega}$ and 2S${\omega}$ was proved by Kamide \cite{KAMIDEBSL06}, with alternative proofs of  cut elimination as consequence of the equivalence theorem. Embedding-based proofs of the cut-elimination and completeness theorems for LT$_{\omega}$ and its propositional fragment were presented by Kamide \cite{KAMIDEMSCS2015}. The present study newly introduces a single-succedent version SLT$_{\omega}$ of LT$_{\omega}$.

Gentzen-style natural deduction systems {\rm PNK} and {\rm PNJ} for classical and intuitionistic until-free LTLs, respectively, were introduced by Baratella and Masini \cite{BM2003}. {\rm PNK} and {\rm PNJ} were regarded as extensions of Gentzen's {\rm NK} and {\rm NJ}, respectively, and were called by the authors the logics of positions. A natural deduction system {\rm PLTL$_{ND}$} was introduced by Bolotov et al. {\rm \cite{BBGS2006}} for a full classical propositional LTL with the until operator {\rm U}. {\rm PLTL$_{ND}$} uses  labelled formulas of the form $i : \al$ and a temporal induction rule concerning the next-time operator $\X$ and the ``globally in the future'' operator $\G$. {\rm PNK}, {\rm PNJ}, and {\rm PLTL$_{ND}$} use an induction rule and do not use infinite premise rules for temporal operators. In contrast, the proposed natural deduction system uses infinite premise rules and do not use an induction rule. By using this setting, we obtain a unified framework.

In this study, we introduce a unified Gentzen-style framework for the until-free propositional logic LTL that  can handle 
Gentzen-style single-succedent sequent calculus and 
 natural deduction 
 uniformly. We  obtain the equivalence among these systems and 
 the fact that  cut elimination for the single-succedent sequent calculus implies  normalization  for the natural deduction system.
 
A unified treatment of the systems of sequent calculus and natural deduction  is the main aim and the original contribution of this study because 
a treatment of this type for 
LTL has not 
been studied to date, instead, sequent calculus and natural deduction for LTL and its fragments have been studied separately. 
A uniform handling of these systems eases the import of meta-results from one formalism to another and is a clear theoretical bonus for their applications.

To address the problem of the correspondence between cut elimination and normalization, we need a Gentzen-style single-succedent sequent calculus because the cut-elimination theorem for usual Gentzen-style multiple-succedent sequent calculi for the standard classical LTL does not imply the normalization theorem for the corresponding natural deduction system. The same situation occurs when considering Gentzen's LK and NK for classical logic.
On the contrary, it is known that  cut elimination  for the single-succedent calculus LJ implies  normalization  for NJ. Thus, we try to obtain an LJ-like single-succedent sequent calculus for the target logic.

To obtain a calculus of this type, we use the following temporal (single-succedent) excluded middle rule:
{\footnotesize
$$
\infer[(\mbox{\rm ex-middle})]{\SEQ{\GA}{\ga}}{
     \SEQ{\X^i\neg\al, \GA}{\ga}
      &
     \SEQ{\X^i\al, \GA}{\ga}
}
$$}where $\X^i$ is an $i$-times nested next-time operator. By using this rule, we can prove the law of excluded middle $\al\LOR\neg\al$. 
The non-temporal version of this rule, which has no occurrence of $\X^i$, was originally introduced by von Plato \cite{von-Plato-draft-1998,NP-2001}. Pursuing  the idea of correspondence between cut elimination and normalization, he introduced a single-succedent sequent calculus for classical logic, 
proved cut elimination, and established normalization  for the corresponding natural deduction system. We thus try to extend this idea to the target temporal logic. Actually, the  single-succedent sequent calculus {\rm SLT$_{\omega}$} proposed in this study can be regarded as a temporal extension of von Plato's calculus and the cut-elimination result for {\rm SLT$_{\omega}$} an extension of his cut-elimination result on classical logic.

Moreover, to obtain the corresponding natural deduction system for the target logic, we use the following rules:
{\footnotesize
$$
\infer[({\rm EXM})]{\ga}{
  \infer*[]{\ga}{
    [\X^i\neg\al]
  }
  &
  \infer*[]{\ga}{
   [\X^i\al]
  }
}
\qquad
\infer[({\rm EXP})]{\ga}{
  \X^i\neg\al
  &
  \X^i\al
}
\qquad
\infer[(\neg {\rm I})]{\X^i\neg\al}{
  \infer*[]{\X^j\neg\ga}{
    [\X^i\al]
  }
  &
  \infer*[]{\X^j\ga}{
   [\X^i\al]
  }
}
$$}where (EXM) corresponds to (ex-middle). 
As mentioned above, the non-temporal version of (EXM), which has no occurrence of $\X^i$, was originally introduced by von Plato \cite{von-Plato-draft-1998,NP-2001} and
the non-temporal version of (EXP) and ($\neg$I) were originally introduced by Gentzen. For more information on these rules, see \cite{elr,saved}. (EXP) has also been used by Bolotov and Shangin \cite{BS2012} for constructing the  paracomplete logic PCont, by K\" urbis and Petrukhin \cite{KP-LLP-2021} for developing some natural deduction systems for a family of many-valued logics including N3, and by Kamide and Negri \cite{KAMIDE-ISMVL-2023-ND,KN-2023} for formalizing Gurevich logic \cite{GUREVICH} and Nelson logic \cite{NELSON,AN84}. Some similar rules to (EXP) were proposed by Priest  \cite{Priest-2019} for constructing natural deduction systems for logics in the FDE (First Degree Entailment) family. (EXP) is regarded as a counterpart rule of (EXM) and is useful for appropriately handling 
natural deduction systems with negation  as a primitive connective (instead of negation defined through implication and the falsity constant). 
The proposed natural deduction system {\rm NLT$_{\omega}$} in this study can thus be regarded as a modified temporal extension of von Plato's classical system 
with the addition of the use of (EXP) and ($\neg$I), and the normalization result for {\rm NLT$_{\omega}$} an extension of the normalization result by von Plato for classical logic. 


\section{Sequent calculus and cut elimination}
\label{sequent-calculi-section}

{\em Formulas} of the logic discussed in this study are constructed using countably many propositional variables, the logical connectives  $\I$ (implication), $\neg$ (negation), $\LAND$ (conjunction), $\LOR$ (disjunction), $\G$ (globally in the future), $\F$ (eventually in the future), and $\X$ (next-time). We use small letters $p, q, ...$ to denote propositional variables and Greek small letters $\al, \be,...$ to denote formulas. 
We use Greek capital letters $\GA, \DE,...$ to denote finite (possibly empty) sets of formulas. 
For any $\M \in \{\G, \F, \X\}$, we use an expression $\M \GA$ to denote the set $\{ \M \ga~|~ \ga \in \GA \}$. The symbol $\equiv$ is used to denote 
definitional equality. The symbol $\omega$ is used to represent the set of natural numbers. An expression $\X^i \al$ for any $i \in \omega$ is defined inductively by $\X^0\al \equiv \al$ and $\X^{n+1}\al \equiv  \X^n \X \al$. We use lower-case letters $i, j$ and $k$ to denote any natural numbers.

We will define Kawai's 
sequent calculus LT$_{\omega}$ \cite{KAWAI87} and a new alternative 
single-succedent sequent calculus SLT$_{\omega}$. Prior to defining these sequent calculi, we need to define some notions and notations.

\begin{df}
A {\em sequent} for {\rm LT$_{\omega}$} is an expression of the form \SEQ{\GA}{\DE}, and a sequent for {\rm SLT$_{\omega}$} is an expression of the form \SEQ{\GA}{\ga} where $\ga$ is a formula or the empty set. 
We use the expression $L \vdash S$ to 
express  the fact that a sequent $S$ is derivable
in a sequent calculus $L$. 
We say that a rule $R$  is {\em admissible} in a sequent calculus $L$ if the following condition is satisfied: For any instance
{\small$
\frac{S_1 \cdots S_n}{S}
$} of $R$, if $L \vdash S_i$ for all $i$, then $L \vdash S$. 
The {\em height} of a derivation in  $L$ is  the number of nodes in a maximal branch of a derivation minus one.  A rule $R$  is {\em height-preserving admissible} if  whenever the premises $S_1 \cdots S_n$ are derivable with height at most $n$ then also the conclusion $S$ is derivable with the same bound on the derivation height.
 Furthermore, we say that $R$ is {\em derivable} in $L$ if there is a derivation in $L$ of $S$ from $S_1, \cdots, S_n$. 
\end{df}



\begin{df}[LT$_{\omega}$]
\label{LT-omega-sequent-calculus-definition}
In the following definitions, $i$ and $k$ represent any natural numbers. 

\noindent The initial sequents of {\rm LT$_{\omega}$} are of the form \SEQ{\X^i p}{\X^i p} for any propositional variable $p$.
\vskip 2pt

\noindent The structural 
rules of {\rm LT$_{\omega}$} are of the form:
{\footnotesize
$$
\infer[({\rm cut})]{\SEQ{\GA,\SI}{\DE,\PI}}{
   \SEQ{\GA}{\DE,\al}
    &
   \SEQ{\al,\SI}{\PI}
}
\quad
\infer[(\mbox{\rm we-left})]{\SEQ{\al,\GA}{\DE}}{
  \SEQ{\GA}{\DE}
}
\quad
\infer[(\mbox{\rm we-right}).]{\SEQ{\GA}{\DE,\al}}{
   \SEQ{\GA}{\DE}
}
$$
}
The logical 
rules of {\rm LT$_{\omega}$} are of the form: 
{\footnotesize
$$
\infer[(\I {\rm left})]{\SEQ{\X^i(\al \I \be), \GA}{\DE}}{
    \SEQ{\GA}{\DE, \X^i\al}
    &
    \SEQ{\X^i\be, \GA}{\DE}
}
\quad
\infer[(\I {\rm right})]{\SEQ{\GA}{\DE, \X^i(\al \I \be)}}{
   \SEQ{\X^i\al, \GA}{\DE, \X^i\be}
}
\quad
\infer[(\neg {\rm left})]{\SEQ{\X^i\neg\al, \GA}{\DE}}{
    \SEQ{\GA}{\DE, \X^i\al}
}
\quad
\infer[(\neg {\rm right})]{\SEQ{\GA}{\DE, \X^i\neg\al}}{
    \SEQ{\X^i\al, \GA}{\DE}
}
$$
$$
\infer[\!\!(\LAND {\rm left})]{\SEQ{\X^i(\al\LAND\be), \GA}{\DE}}{
   \SEQ{\X^i\al, \X^i\be, \GA}{\DE}
}
\quad\!\!
\infer[\!\!(\LAND {\rm right})]{\SEQ{\GA}{\DE, \X^i(\al\LAND\be)}}{
   \SEQ{\GA}{\DE, \X^i\al}
    &
   \SEQ{\GA}{\DE, \X^i\be}
}
\quad\!\!
\infer[\!\!(\LOR {\rm left})]{\SEQ{\X^i(\al\LOR\be), \GA}{\DE}}{
      \SEQ{\X^i\al, \GA}{\DE}
       &
      \SEQ{\X^i\be, \GA}{\DE}
}
\quad\!\!
\infer[\!\!(\LOR {\rm right})]{\SEQ{\GA}{\DE, \X^i(\al\LOR\be)}}{
    \SEQ{\GA}{\DE, \X^i\al, \X^i\be}
} 
$$
$$
\infer[(\G {\rm left})]{\SEQ{\X^i \G \al, \GA}{\DE}}{
    \SEQ{\X^{i+k} \al, \GA}{\DE}
}
\quad 
\infer[(\G {\rm right})]{\SEQ{\GA}{\DE, \X^i \G \al}}{
    \{~ \SEQ{\GA}{\DE, \X^{i+j} \al}~ \}_{j \in \omega} 
}
\quad
\infer[(\F {\rm left})]{\SEQ{\X^i \F \al, \GA}{\DE}}{
    \{~ \SEQ{\X^{i+j} \al, \GA}{\DE}~ \}_{j \in \omega} 
}
\quad 
\infer[(\F {\rm right}).]{\SEQ{\GA}{\DE, \X^i \F \al}}{
    \SEQ{\GA}{\DE, \X^{i+k} \al}
}
$$
}
\end{df}

\begin{rmk}~
The calculus {\rm LT$_{\omega}$} introduced here is a slightly modified propositional version of Kawai's sequent calculus {\rm \cite{KAWAI87}} for until-free first-order linear-time temporal logic. 
The following cut-elimination theorem holds for {\rm LT$_{\omega}$}. 
The rule {\rm (cut)} is admissible in cut-free {\rm LT$_{\omega}$}. 
We will use this theorem in the following discussion. 
The cut-elimination theorem for (the original first-order) {\rm LT$_{\omega}$} was proved by Kawai in {\rm \cite{KAWAI87}}.  
\end{rmk}


Next, we introduce SLT$_{\omega}$. We use the same names for the 
rules of SLT$_{\omega}$ as those of LT$_{\omega}$, although the forms of the rules are different.

\begin{df}[SLT$_{\omega}$]
In the following definitions, $i$ and $k$ represent any natural numbers and $\ga$ represents a formula or the empty set.

\noindent The initial sequents of {\rm SLT$_{\omega}$} are of the form 
\SEQ{\X^i p,\Gamma}{\X^i p}
for any propositional variable $p$.

\vskip 2pt

\noindent The structural 
rules of {\rm SLT$_{\omega}$} are of the form:
{\footnotesize
$$
\infer[({\rm cut})]{\SEQ{\GA,\SI}{\ga}}{
   \SEQ{\GA}{\al}
    &
   \SEQ{\al,\SI}{\ga}
}
\quad
\infer[(\mbox{\rm we-right}).]{\SEQ{\GA}{\al}}{
   \SEQ{\GA}{}
}
$$
}
The logical 
rules of {\rm SLT$_{\omega}$} are of the form: 
{\footnotesize
$$ 
\infer[(\I {\rm left})]{\SEQ{\X^i(\al \I \be),\GA}{\ga}}{
    \SEQ{\GA}{\X^i\al}
    &
    \SEQ{\X^i\be, \GA}{\ga}
}
\quad 
\infer[(\I {\rm right})]{\SEQ{\GA}{\X^i(\al \I \be)}}{
   \SEQ{\X^i\al, \GA}{\X^i\be}
}
\quad
\infer[(\neg {\rm left})]{\SEQ{\X^i\neg\al, \GA}{}}{
   \SEQ{\GA}{\X^i\al}
}
\quad 
\infer[(\neg {\rm right})]{\SEQ{\GA}{\X^i\neg\al}}{
   \SEQ{\X^i\al, \GA}{}
}
$$
$$
\infer[(\mbox{\rm ex-middle})]{\SEQ{\GA}{\ga}}{
     \SEQ{\X^i\neg\al, \GA}{\ga}
      &
     \SEQ{\X^i\al, \GA}{\ga}
}
\quad
\infer[(\land {\rm left})]{\SEQ{\X^i(\al\land\be), \GA}{\ga}}{
   \SEQ{\X^i\al, \X^i\be, \GA}{\ga}
}
\quad 
\infer[(\land {\rm right})]{\SEQ{\GA}{\X^i(\al\land\be)}}{
   \SEQ{\GA}{\X^i\al}
    &
   \SEQ{\GA}{\X^i\be}
}
$$
$$
\infer[( \lor {\rm left})]{\SEQ{\X^i(\al\lor\be),\GA}{\ga}}{
      \SEQ{\X^i\al, \GA}{\ga}
       &
      \SEQ{\X^i\be, \GA}{\ga}
}
\quad
\infer[(\lor {\rm right1})]{\SEQ{\GA}{\X^i(\al\lor\be)}}{
    \SEQ{\GA}{\X^i\al}
}
\quad 
\infer[(\lor {\rm right2})]{\SEQ{\GA}{\X^i(\al\lor\be)}}{
    \SEQ{\GA}{\X^i\be}
}
$$
$$
\infer[(\G {\rm left})]{\SEQ{\X^i \G \al, \GA}{\ga}}{
    \SEQ{\X^{i+k} \al, \GA}{\ga}
}
\quad
\infer[(\G {\rm right})]{\SEQ{\GA}{\X^i \G \al}}{
    \{~ \SEQ{\GA}{\X^{i+j} \al}~ \}_{j \in \omega} 
}
\quad
\infer[(\F {\rm left})]{\SEQ{\X^i \F \al, \GA}{\ga}}{
    \{~ \SEQ{\X^{i+j} \al, \GA}{\ga}~ \}_{j \in \omega} 
}
\quad
\infer[(\F {\rm right}).]{\SEQ{\GA}{\X^i \F \al}}{
    \SEQ{\GA}{\X^{i+k} \al}
}
$$
}
\end{df}

\begin{prop}
\label{slt-prop-1}
Let $L$ be  {\rm LT$_{\omega}$} or {\rm SLT$_{\omega}$}.   
The sequents of the form 
\SEQ{\X^i\al, \GA}{\X^i\al} 
for any formula $\al$ and any natural number $i$ are 
derivable
in $L$.
\end{prop}
\PROOF
By induction on $\al$.
\QED

\begin{prop}
\label{we-left-adm}
The 
following rule is height-preserving admissible in cut-free {\rm SLT$_{\omega}$}: 
{\footnotesize
$$
\infer[(\mbox{\rm we-left}).]{\SEQ{\al,\GA}{\ga}}{
  \SEQ{\GA}{\ga}
}
$$
}
\end{prop}
\PROOF
By straightforward induction on the height of the derivation since weakening is in-built in initial sequents and all the rules have an arbitrary context on the left.
\QED
\\


Next, we show the cut-elimination theorem for SLT$_{\omega}$ using the method by Africk \cite{AFRICK}. We also prove a theorem that 
establishes an equivalence between SLT$_{\omega}$ and LT$_{\omega}$.  
Prior to proving these theorems, we show the following proposition and lemmas.

\begin{prop}
\label{SLT-omega-prop}
The following rule is derivable in cut-free {\rm SLT$_{\omega}$}:
{\footnotesize
$$
\infer[(\neg{\rm left}^{-1}).]{\SEQ{\GA}{\X^i\al}}{
      \SEQ{\X^i\neg\al, \GA}{}
}
$$
}
\end{prop}
\PROOF
By using (ex-middle), 
(we-right), and Proposition \ref{slt-prop-1}.
\QED

\begin{lm}
\label{SLT-omega-lemma}
For any sequent \SEQ{\GA}{\DE}, 
if {\rm LT$_{\omega}$} $-$ {\rm (cut)} $\vdash$ \SEQ{\GA}{\DE}, then 
{\rm SLT$_{\omega}$} $-$ {\rm (cut)} $\vdash$ \SEQ{\neg\DE, \GA}{}. 
\end{lm}
\PROOF
 By induction on the derivations ${\mathcal{D}}$ of \SEQ{\GA}{\DE} in cut-free LT$_{\omega}$. 
We distinguish the cases according to the last inference 
 of ${\mathcal{D}}$. 
We show only the case of ($\LOR$right) as follows. 
The last inference of ${\mathcal{D}}$ is of the form:
{\footnotesize
$$
\infer[(\LOR {\rm right}).]{\SEQ{\GA}{\DE, \X^i(\al\LOR\be)}}{
    \SEQ{\GA}{\DE, \X^i\al, \X^i\be}
}
$$}By induction hypothesis, we have SLT$_{\omega}$ $-$ (cut) $\vdash$ \SEQ{\neg\X^i\al, \neg\X^i\be, \neg\DE, \GA}{}.
Then, we obtain the required derivation:
{\footnotesize
$$
\infer[(\neg {\rm left})]{\SEQ{\neg\X^i(\al\LOR\be), \neg\X^i(\al\LOR\be), \neg\DE,\GA}{}}{
\infer[(\LOR {\rm right2})]{\SEQ{\neg\X^i(\al\LOR\be), \neg\DE, \GA}{\X^i(\al\LOR\be)}}{
\infer[(\neg {\rm left}^{-1})]{\SEQ{\neg\X^i(\al\LOR\be), \neg\DE, \GA}{\X^i\be}}{
\infer[(\neg {\rm left})]{\SEQ{\neg\X^i(\al\LOR\be), \neg\X^i\be, \neg\DE, \GA}{}}{
     \infer[(\LOR {\rm right1})]{\SEQ{\neg\X^i\be, \neg\DE, \GA}{\X^i (\al\LOR\be)}}{
          \infer[(\neg {\rm left}^{-1})]{\SEQ{\neg\X^i\be, \neg\DE, \GA}{\X^i\al}}{
                \infer*[Ind.\, hyp.]{\SEQ{\neg\X^i\al, \neg\X^i\be, \neg\DE, \GA}{}}{
                }
           }
     }
}
}
}
}
$$}where 
\SEQ{\neg\X^i(\al\LOR\be), \neg\X^i(\al\LOR\be), \neg\DE, \GA}{} is equivalent to \SEQ{\neg\X^i(\al\LOR\be), \neg\DE, \GA}{} (because the antecedent of the sequent is a set of formulas) and ($\neg$left$^{-1}$) is derivable in cut-free SLT$_{\omega}$ by Proposition \ref{SLT-omega-prop}. 
\QED

\begin{lm}
\label{SLT-omega-lemma-2}
For any sequent \SEQ{\GA}{\ga}, 
if {\rm SLT$_{\omega}$} $\vdash$ \SEQ{\GA}{\ga}, then 
{\rm LT$_{\omega}$} $\vdash$ \SEQ{\GA}{\ga}. 
\end{lm}
\PROOF
By induction on the derivations ${\mathcal{D}}$ of  \SEQ{\GA}{\ga} in SLT$_{\omega}$. 
We distinguish the cases according to the last inference of ${\mathcal{D}}$.
An initial sequent of  {\rm SLT$_{\omega}$}, i.e. of the form \SEQ{\X^i p,\Gamma}{\X^i p}, 
is derived from an initial sequent of  {\rm LT$_{\omega}$} using weakening steps.
Next, we show only the critical case of (ex-middle) as follows.
The last inference of ${\mathcal{D}}$ is fo the form:
{\footnotesize
$$
\infer[(\mbox{ex-middle}).]{\SEQ{\GA}{\ga}}{
      \infer*[]{\SEQ{\X^i \neg\al, \GA}{\ga}}{
       }
       &
       \infer*[]{\SEQ{\X^i \al, \GA}{\ga}}{
       }
}
$$}By induction hypotheses, we have
LT$_{\omega}$ $\vdash$ \SEQ{\X^i\neg\al, \GA}{\ga} and 
LT$_{\omega}$ $\vdash$ \SEQ{\X^i\al, \GA}{\ga}.
We then obtain the required derivation: 
{\footnotesize
$$
\infer[({\rm cut})]{\SEQ{\GA}{\ga}}{
      \infer[({\rm cut})]{\SEQ{\GA}{\ga, \X^i\al}}{
             \infer[(\neg {\rm right})]{\SEQ{}{\X^i\al, \X^i\neg\al}}{
                   \infer*[Prop.~\ref{slt-prop-1}]{\SEQ{\X^i\al}{\X^i\al}}{
                   }
              }
              &
              \infer*[Ind.\, hyp.]{\SEQ{\X^i\neg\al, \GA}{\ga}}{
              }
      }
      &
      \infer*[Ind.\, hyp.]{\SEQ{\X^i\al, \GA}{\ga}}{
      }
}
$$}
\QED

\begin{thm}[Cut elimination for SLT$_{\omega}$]
\label{cut-eli-SLT-omega}
The rule {\rm (cut)} is admissible in cut-free {\rm SLT$_{\omega}$}.
\end{thm}
\PROOF
Suppose SLT$_{\omega}$ $\vdash$ \SEQ{\GA}{\ga}. 
Then, we obtain LT$_{\omega}$ $\vdash$ \SEQ{\GA}{\ga} by Lemma \ref{SLT-omega-lemma-2}. 
Thus,  we have LT$_{\omega}$ $-$ (cut) $\vdash$ \SEQ{\GA}{\ga} by the cut-elimination theorem for LT$_{\omega}$ \cite{KAWAI87,KAMIDEBSL06}. 
Thus, we obtain SLT$_{\omega}$ $-$ (cut) $\vdash$ \SEQ{\neg\ga, \GA}{} by Lemma \ref{SLT-omega-lemma}. 
We thus obtain the required fact SLT$_{\omega}$ $-$ (cut) $\vdash$ \SEQ{\GA}{\ga} by applying ($\neg$left$^{-1}$) to \SEQ{\neg\ga, \GA}{}, where ($\neg$left$^{-1}$) is derivable in cut-free SLT$_{\omega}$ by Proposition \ref{SLT-omega-prop}. 
\QED

\begin{thm}[Equivalence between SLT$_{\omega}$ and LT$_{\omega}$]
For any formula $\al$, 
{\rm SLT$_{\omega}$} $\vdash$ \SEQ{}{\al} iff  
{\rm LT$_{\omega}$} $\vdash$ \SEQ{}{\al}. 
\end{thm}
\PROOF
($\Longrightarrow$):  
By Lemma \ref{SLT-omega-lemma-2}.
($\Longleftarrow$): 
Suppose LT$_{\omega}$ $\vdash$ \SEQ{}{\al}. 
Then, we obtain LT$_{\omega}$ $-$ (cut) $\vdash$ \SEQ{}{\al} by the cut-elimination theorem for LT$_{\omega}$ \cite{KAWAI87,KAMIDEBSL06}.
We then obtain SLT$_{\omega}$ $-$ (cut) $\vdash$ \SEQ{\neg\al}{} by Lemma \ref{SLT-omega-lemma}.
Thus, we obtain the required fact SLT$_{\omega}$ $\vdash$ \SEQ{}{\al} by applying ($\neg$left$^{-1}$) to \SEQ{\neg\al}{}, where ($\neg$left$^{-1}$) is derivable in cut-free SLT$_{\omega}$ by Proposition \ref{SLT-omega-prop}. 
\QED

\section{Natural deduction}
\label{natural-deduction-section}


As usual in the definition of a natural deduction system, the notation $[\al]$  denotes that the formula $\al$ is a discharged assumption by the underlying logical inference rule.

We define a Gentzen-style natural deduction system NLT$_{\omega}$ for until-free propositional LTL.


\begin{df}[NLT$_{\omega}$]
\label{nlt-omega}
Let $i$ and $k$ be any natural numbers. 
The logical 
rules of {\rm NLT$_{\omega}$} are of the following form, 
where in {\rm ($\I$I)} the discharge can be vacuous: 
{\footnotesize
$$
\infer[(\I {\rm I})]{\X^i (\al\I\be)}{
  \infer*[]{\X^i\be}{
    [\X^i\al]
  }
}
\quad
\infer[(\I {\rm E})]{\X^i\be}{
  \X^i (\al\I\be)
  &
  \X^i \al
}
\quad
\infer[({\rm EXP})]{\ga}{
  \X^i\neg\al
  &
  \X^i\al
}
\quad
\infer[({\rm EXM})]{\ga}{
  \infer*[]{\ga}{
    [\X^i\neg\al]
  }
  &
  \infer*[]{\ga}{
   [\X^i\al]
  }
}
$$
$$
\infer[(\neg {\rm I})]{\X^i\neg\al}{
  \infer*[]{\X^j\neg\ga}{
    [\X^i\al]
  }
  &
  \infer*[]{\X^j\ga}{
   [\X^i\al]
  }
}
\quad
\infer[(\LAND {\rm I})]{\X^i (\al\LAND\be)}{
     \X^i\al
     &
     \X^i\be
}
\quad
\infer[(\LAND {\rm E1})]{\X^i\al}{
     \X^i (\al\LAND\be)
}
\quad
\infer[(\LAND {\rm E2})]{\X^i\be}{
     \X^i (\al\LAND\be)
}
$$
$$
\infer[(\LOR {\rm I1})]{\X^i (\al\LOR \be)}{
    \X^i\al
}
\quad
\infer[(\LOR {\rm I2})]{\X^i (\al\LOR \be)}{
  \X^i\be
}
\quad
\infer[(\LOR {\rm E})]{\ga}{
   \X^i (\al\LOR\be)
  &
  \infer*[]{\ga}{
     [\X^i\al]
   }
   &
  \infer*[]{\ga}{
     [\X^i\be]
  }
}
$$
$$
\infer[(\G {\rm I})]{\X^i \G \al}{
\{~ \X^{i+j}\al~\}_{j \in \omega} 
}
\quad
\infer[(\G {\rm E})]{\X^{i+k} \al}{
    \X^i \G \al
}
\quad
\infer[(\F {\rm I})]{\X^i \F\al}{
   \X^{i+k}\al
}
\quad
\infer[(\F {\rm E}).]{\ga}{
     \X^i \F \al
     &
     \infer*[]{\{~\ga~\}_{j\in \omega}}{
         [\X^{i+j}\al]
     }
}
$$
}
\end{df}

\begin{rmk}
{\rm (EXP)}, {\rm (EXM)}, and {\rm ($\neg$I)} are characteristic rules in {\rm NLT$_{\omega}$}. The rule {\rm (EXP)} and {\rm ($\neg$I)} are temporal generalizations of the original rules introduced by Gentzen. The rule {\rm (EXM)} is a temporal generalization of the original rule introduced by von Plato {\rm \cite{von-Plato-draft-1998,NP-2001}}. 
The non-temporal versions of {\rm (EXP)}, {\rm (EXM)}, and {\rm ($\neg$I)} were also used by Kamide and Negri in {\rm \cite{KN-2023}} for constructing natural deduction systems for logics with strong negation. 
Using {\rm (EXP)} and {\rm (EXM)}, we can prove the formulas of the form $(\neg\al\LAND\al)\I\ga$ and $\neg\al\LOR\al$, respectively. 
Using {\rm ($\neg$I)} and {\rm (EXP)}, we can prove the formulas of the form $\al\I\neg\neg\al$ and $\neg\neg(\al\I\al)$ by:
{\footnotesize
$$
\infer[(\I {\rm I})^1]{\al\I\neg\neg\al}{
     \infer[(\neg {\rm I})^2]{\neg\neg\al }{
           \infer[({\rm EXP})]{\neg\al}{
                  [\neg\al]^2
                   &
                  [\al]^1
           }
           &
           \infer[({\rm EXP})]{\al}{
                  [\neg\al]^2
                   &
                  [\al]^1
           }
     }
}
\quad
\infer[(\neg {\rm I})^1.]{\neg\neg (\al\I\al)}{
     \infer[({\rm EXP})]{\al\I\al}{
          \infer[(\I {\rm I})^3]{\al\I\al}{
               [\al]^3
          }
         &
                [\neg (\al\I\al)]^1
    }
     &
     \infer[({\rm EXP})]{\neg (\al\I\al)}{
            \infer[(\I {\rm I})^2]{\al\I\al}{
               [\al]^2
          }
          &
                [\neg (\al\I\al)]^1
     }
}
$$
}
\end{rmk}

Next, we define some notions for NLT$_{\omega}$.

\begin{df}
The 
rules 
{\rm ($\I$I)}, 
{\rm ($\LAND$I)}, {\rm ($\LOR$I1)}, {\rm ($\LOR$I2)}, {\rm ($\neg$I)}, {\rm ($\G$I)}, {\rm ($\F$I)}, and {\rm (EXM)} are called {\em introduction rules}, and the 
rules {\rm ($\I$E)}, {\rm ($\LAND$E1)}, {\rm ($\LAND$E2)}, {\rm ($\LOR$E)}, {\rm ($\G$E)}, {\rm ($\F$E)}, and {\rm (EXP)} are called {\em elimination rules}. 
The notions of {\em major} and {\rm minor premises} of the 
rules without {\rm (EXM)} and  {\rm (EXP)} are defined as usual. If $\X^i\neg\al$ and $\X^i\al$ are both premises of {\rm (EXP)}, then $\X^i\neg\al$ and $\X^i\al$ are called the major and minor premises of {\rm (EXP)}, respectively. 
The notions of {\em derivation},
{\em (open and discharged) assumptions} of a derivation, and {\em end-formula} of a derivation are also defined as usual. 
For a derivation ${\mathcal{D}}$, we use the expression {\rm oa(${\mathcal{D}}$)} to denote the
set 
of open assumptions of ${\mathcal{D}}$ and the expression {\rm end(${\mathcal{D}}$)} to denote the end-formula of ${\mathcal{D}}$. 
A formula $\al$ is said to be {\em provable} in a natural deduction system $L$ if there exists a derivation of $L$ with no open assumption whose end-formula is $\al$. 
\end{df}

\begin{rmk}
There are no notions of  major and minor premises of {\rm (EXM)} and {\rm ($\neg$I)}. Namely, the premises of {\rm (EXM)} and {\rm ($\neg$I)} are neither major nor minor premises. 
In this study, {\rm (EXP)} is treated as an elimination rule, and {\rm (EXM)} is treated as an introduction rule. 
\end{rmk}

Next, we define a reduction relation $\gg$ on the set of derivations in NLT$_{\omega}$. 
Prior to defining $\gg$, we define some notions concerning $\gg$.

\begin{df}
Let $\al$ be a formula occurring in a derivation ${\mathcal{D}}$ in {\rm NLT$_{\omega}$}. 
Then, $\al$ is called a {\em maximum formula} in ${\mathcal{D}}$ if $\al$ satisfies the following conditions: 
(1) $\al$ is the conclusion of an introduction rule, {\rm ($\LOR$E)}, or {\rm (EXP)} and 
%
(2) $\al$ is the major premise of an elimination rule. 
A derivation is said to be {\em normal} if it contains no maximum formula. 
The notion of substitution of derivations for assumptions is defined as usual. We assume that the set of derivations is closed under substitution. 
\end{df}

\begin{df}[Reduction relation]
\label{nlt-omega-reduction}
Let $\ga$ be a maximum formula in a derivation that is the conclusion of a
rule $R$. 
The definition of the reduction relation $\gg$ at $\ga$ in {\rm NLT$_{\omega}$} is obtained by the following conditions.
\begin{enumerate}
\item
$R$ is 
{\rm ($\I$I)} 
and $\ga$ is $\X^i (\al\I\be)$: 
{\footnotesize
$$
\infer[(\I {\rm E})]{\X^i \be}{
   \infer[(\I {\rm I})]{\X^i (\al\I\be)}{
      \infer*[{\mathcal{D}}]{\X^i \be}{
         [\X^i \al]
      }
   }
   &
   \infer*[{\mathcal{E}}]{\X^i\al}{
   }
}
\quad\quad \gg \quad\quad
  \infer*[{\mathcal{D}}]{\X^i\be.}{
     \infer*[{\mathcal{E}}]{\X^i\al}{
      }
  }
$$
}
\item
$R$ is {\rm (EXP)}: 
{\footnotesize
$$
\infer[R']{\pi}{
  \infer[({\rm EXP})]{\ga}{
      \infer*[{\mathcal{D}}_1]{\X^i\neg\de}{
       }
      &
       \infer*[{\mathcal{D}}_2]{\X^i\de}{
       }
  }
  &
  \infer*[{\mathcal{E}}_1]{\pi_1}{
  }
  &
  \infer*[{\mathcal{E}}_2]{\pi_2}{
  }
}
\quad\quad \gg \quad\quad 
\infer[({\rm EXP})]{\pi}{
      \infer*[{\mathcal{D}}_1]{\X^i\neg\de}{
  }
  &
      \infer*[{\mathcal{D}}_2]{\X^i\de}{
   }
}
$$}where $R'$ is an arbitrary 
rule, and both ${\mathcal{E}}_1$ and ${\mathcal{E}}_2$ are derivations of the minor premises of $R'$ if they exist. 
\item 
$R$ is {\rm ($\neg$I)}, $\ga$ is $\X^i\neg\al$, and $\be$ is the conclusion of {\rm (EXP)}:
{\footnotesize
$$
\infer[({\rm EXP})]{\be}{
  \infer[(\neg {\rm I})]{\X^i\neg\al}{
      \infer*[{\mathcal{D}}_1]{\X^j\neg\de}{
           [\X^i\al]
       }
      &
       \infer*[{\mathcal{D}}_2]{\X^j\de}{
          [\X^i\al]
       }
  }
  &
  \infer*[{\mathcal{E}}]{\X^i\al}{
  }
}
\quad\quad \gg \quad\quad 
\infer[({\rm EXP}).]{\be}{
      \infer*[{\mathcal{D}}_1]{\X^j\neg\de}{
          \infer*[{\mathcal{E}}]{\X^i\al}{}
  }
  &
      \infer*[{\mathcal{D}}_2]{\X^j\de}{
          \infer*[{\mathcal{E}}]{\X^i\al}{}
   }
}
$$
}
\item 
$R$ is {\rm ($\neg$I)}, $\ga$ is $\X^i\neg\de$, 
and $\X^i\de$ is the conclusion of {\rm (EXP)}:
{\footnotesize
$$
\infer[({\rm EXP})]{\X^i\de}{
  \infer[(\neg {\rm I})]{\X^i\neg\de}{
      \infer*[{\mathcal{D}}_1]{\X^j\neg\be}{
           [\X^i\de]
       }
      &
       \infer*[{\mathcal{D}}_2]{\X^j\be}{
          [\X^i\de]
       }
  }
  &
  \infer*[{\mathcal{E}}]{\X^i\de}{
  }
}
\quad\quad \gg \quad\quad 
\infer*[{\mathcal{E}}]{\X^i\de}{
}
$$
}
\item
$R$ is {\rm (EXM)} and $\ga$ is $\X^i(\ga_1\I\ga_2)$, $\X^i(\ga_1\LAND\ga_2)$, or $\X^i (\ga_1\LOR\ga_2)$: 
{\footnotesize
$$
\infer[R']{\de}{
   \infer[({\rm EXM})]{\ga}{
       \infer*[{\mathcal D}_1]{\ga}{
          [\X^i\neg\al]
        }
        &
       \infer*[{\mathcal D}_2]{\ga}{
          [\X^i\al]
        }
   }  
   &
   \infer*[{\mathcal E}_1]{\de_1}{
   }
   &
   \infer*[{\mathcal E}_2]{\de_2}{
   }
}
\quad\quad \gg \quad\quad 
\infer[({\rm EXM})]{\de}{ 
   \infer[R']{\de}{
      \infer*[{\mathcal D}_1]{\ga}{
          [\X^i\neg\al]
       }
       &
       \infer*[{\mathcal E}_1]{\de_1}{
       }
       &
       \infer*[{\mathcal E}_2]{\de_2}{
       }
   }
   &
   \infer[R']{\de}{
      \infer*[{\mathcal D}_2]{\ga}{
          [\X^i\al]
       }
       &
       \infer*[{\mathcal E}_1]{\de_1}{
       }
       &
       \infer*[{\mathcal E}_2]{\de_2}{
       }
   }
}
$$}where $R'$ is 
{\rm ($\I$E)}, 
{\rm ($\LAND$E1)}, 
{\rm ($\LAND$E2)}, or
{\rm ($\LOR$E)},
and both ${\mathcal E}_1$ and ${\mathcal E}_2$ are derivations of the minor premises of $R'$ if they exist. 
\item
$R$ is {\rm (EXM)}, $\ga$ is $\X^i\neg\de$, and $\X^i\de$ is the conclusion of {\rm (EXP)}:  
{\footnotesize
$$
\infer[{\rm (EXP)}]{\X^i\de}{
  \infer[({\rm EXM})]{\X^i\neg\de}{
      \infer*[{\mathcal{D}}_1]{\X^i\neg\de}{
         [\X^i\neg\al]
       }
      &
       \infer*[{\mathcal{D}}_2]{\X^i\neg\de}{
         [\X^i\al]
       }
  }
  &
  \infer*[{\mathcal{E}}]{\X^i\de}{
  }
}
\quad\quad \gg \quad\quad
\infer*[{\mathcal{E}}]{\X^i\de.}{
}
$$
}
\item
$R$ is {\rm ($\LAND$I)} and $\ga$ is $\X^i (\al_1\LAND\al_2)$: 
{\footnotesize
$$
\infer[(\LAND {\rm E}i)]{\X^i\al_i}{
   \infer[(\LAND {\rm I})]{\X^i (\al_1\LAND\al_2)}{
       \infer*[{\mathcal{D}}_1]{\X^i \al_1}{
        }
        &
       \infer*[{\mathcal{D}}_2]{\X^i \al_2}{
        }
   }  
}
\quad\quad \gg \quad\quad 
   \infer*[{\mathcal{D}}_i]{\X^i \al_i}{
   }
\quad where~i~is~1~or~2. 
$$}
\item
$R$ is {\rm ($\LOR$I1)} or {\rm ($\LOR$I2)} and $\ga$ is $\X^i (\al_1\LOR\al_2)$: 
{\footnotesize
$$
\infer[(\LOR {\rm E})]{\de}{
  \infer[(\LOR {\rm I}i)]{\X^i (\al_1\LOR\al_2)}{
      \infer*[{\mathcal{D}}]{\X^i \al_i}{
       }
  }
  &
  \infer*[{\mathcal{E}}_1]{\de}{
    [\X^i\al_1]
  }
  &
  \infer*[{\mathcal{E}}_2]{\de}{
    [\X^i\al_2]
  }
}
\quad\quad \gg \quad\quad 
  \infer*[{\mathcal{E}}_i]{\de}{
      \infer*[{\mathcal{D}}]{\X^i\al_i}{
      }
  }
\quad where~i~is~1~or~2. 
$$}
\item
$R$ is {\rm ($\LOR$E)}: 
{\footnotesize
$$
\infer[R']{\de}{
   \infer[(\LOR {\rm E})]{\pi}{
       \infer*[{\mathcal{D}}_1]{\X^i (\al\LOR\be)}{
        }
        &
       \infer*[{\mathcal{D}}_2]{\pi}{
          [\X^i\al]
        }
        &
       \infer*[{\mathcal{D}}_3]{\pi}{
          [\X^i\be]
        }
   }  
   &
   \infer*[{\mathcal{E}}_n]{\{~\de_n~\}}{
   }
}
\quad\quad \gg \quad\quad 
\infer[(\LOR {\rm E})]{\de}{ 
   \infer*[{\mathcal{D}}_1]{\X^i (\al\LOR\be)}{
   }
   &
   \infer[R']{\de}{
      \infer*[{\mathcal{D}}_2]{\pi}{
          [\X^i\al]
       }
       &
       \infer*[{\mathcal{E}}_n]{\{~\de_n~\}}{
       }
   }
   &
   \infer[R']{\de}{
      \infer*[{\mathcal{D}}_3]{\pi}{
          [\X^i\be]
       }
       &
       \infer*[{\mathcal{E}}_n]{\{~\de_n~\}}{
       }
   }
}
$$}where $R'$ is an arbitrary 
rule, and ${\mathcal{E}}_1$, ${\mathcal{E}}_2$, ... , ${\mathcal{E}}_n$, ... are derivations of the minor premises of $R'$ if they exist. 
\item
$R$ is {\rm ($\G$I)} and $\ga$ is $\X^i\G \al$: 
{\footnotesize
$$
\infer[(\G {\rm E})]{\X^{i+k} \al}{
   \infer[(\G {\rm I})]{\X^i \G \al}{
       \infer*[{\mathcal{D}}_j]{\{~ \X^{i+j} \al~\}_{j\in \omega}}{
        }
   }  
}
\quad\quad \gg \quad\quad 
   \infer*[{\mathcal{D}}_k]{\X^{i+k} \al}{
   }
\quad where~k \in \omega.
$$}
\item
$R$ is {\rm ($\F$I)} and $\ga$ is $\X^i \F\al$: 
{\footnotesize
$$
\infer[(\F {\rm E})]{\de}{
  \infer[(\F {\rm I})]{\X^i \F\al}{
      \infer*[{\mathcal{D}}_k]{\X^{i+k} \al}{
       }
  }
  &
  \infer*[{\mathcal{E}}_j]{\{~\de~\}_{j\in \omega}}{
    [\X^{i+j} \al]
  }
}
\quad\quad \gg \quad\quad 
  \infer*[{\mathcal{E}}_k]{\de}{
      \infer*[{\mathcal{D}}_k]{\X^{i+k}\al}{
      }
  }
\quad where~k \in\omega. 
$$}
\item
$R$ is {\rm ($\F$E)}: 
{\footnotesize
$$
\infer[R']{\de}{
   \infer[(\F {\rm E})]{\pi}{
       \infer*[{\mathcal{D}}]{\X^i \F\al}{
        }
        &
       \infer*[{\mathcal{D}}_j]{\{~\pi~\}_{j\in\omega}}{
          [\X^{i+j}\al]
        }
   }  
   &
   \infer*[{\mathcal{E}}_n]{\{~\de_n~\}}{
   }
}
\quad\quad \gg \quad\quad 
\infer[(\F {\rm E})]{\de}{ 
   \infer*[{\mathcal{D}}]{\X^i \F\al}{
   }
   &
   \infer[R']{\{~\de~\}_{j\in\omega}}{
      \infer*[{\mathcal{D}}_j]{\pi}{
          [\X^{i+j}\al]
       }
       &
       \infer*[{\mathcal{E}}_n]{\{~\de_n~\}}{
       }
   }
}
$$}where $R'$ is an arbitrary 
rule, and ${\mathcal{E}}_1$, ${\mathcal{E}}_2$, ... , ${\mathcal{E}}_n$, ... are derivations of the minor premises of $R'$ if they exist. 
\item
The set of derivations are closed under $\gg$. 
\end{enumerate}
\end{df}


\begin{df}
If ${\mathcal{D}}'$ is obtained from ${\mathcal{D}}$ by the reduction relation of 
Definition \ref{nlt-omega-reduction}, we write
 ${\mathcal{D}} \gg {\mathcal{D}}'$. A sequence ${\mathcal{D}}_0, {\mathcal{D}}_1, ...$ of derivations is called a {\em reduction sequence} if it satisfies the following conditions: 
(1) ${\mathcal{D}}_i \gg {\mathcal{D}}_{i+1}$ for all $i \geq 0$, and 
(2) the last derivation in the sequence is normal if the sequence is finite. 
A derivation ${\mathcal{D}}$ is called {\em normalizable} if there is a finite reduction sequence starting from ${\mathcal{D}}$. 
\end{df}

\section{Equivalence and normalization}
\label{equivalence-normalization-section}

In the following discussion, a derivation of \SEQ{\GA}{} in SLT$_{\omega}$ is interpreted as a derivation ${\mathcal{D}}$ in NLT$_{\omega}$ such that oa(${\mathcal{D}}$) = $\GA$ and end(${\mathcal{D}}$) = $\neg p\LAND p$.


\begin{lm}
\label{prop-nlt-slt}
We have the following statements.  
\begin{enumerate}
\item 
If ${\mathcal{D}}$ is a derivation in {\rm NLT$_{\omega}$} such that 
{\rm oa(${\mathcal{D}}$)} $=$ $\GA$ and 
{\rm end(${\mathcal{D}}$)} $=$ $\be$, 
then {\rm SLT$_{\omega}$} $\vdash$ \SEQ{\GA}{\be},

\item 
If {\rm SLT$_{\omega}$} $-$ {\rm (cut)} $\vdash$ \SEQ{\GA}{\be},  
then we  obtain a derivation ${\mathcal{D}}'$ in {\rm NLT$_{\omega}$} such that 
(a) {\rm oa(${\mathcal{D}}'$)} $=$ $\GA$, 
%
(b) {\rm end(${\mathcal{D}}'$)} $=$ $\be$, and 
%
(c) ${\mathcal{D}}'$ is normal.
\end{enumerate}
\end{lm}
\PROOF
\begin{enumerate}
\item
We prove 1 by induction on the derivations ${\mathcal{D}}$ of NLT$_{\omega}$ such that oa(${\mathcal{D}}$) = $\GA$ and end(${\mathcal{D}}$) = $\be$. We distinguish the cases according to the last inference of ${\mathcal{D}}$. We show some cases. 
Observe that we shall use (we-left), which  is admissible by Proposition \ref{we-left-adm}.

\begin{enumerate}
\item
Case ($\I$I): 
We show only the following subcase, which has no discharged assumption $[\X^i\al]$. 
${\mathcal{D}}$ is of the form: 
{\footnotesize
$$
\infer[{\rm (\I I)}]{\X^i(\al\I\ga)}{
     \infer*[{\mathcal{E}}]{\X^i\ga}{
          \GA
     }
}
$$}where oa(${\mathcal{D}}$) = $\GA$ and end(${\mathcal{D}}$) = $\ga$. 
By induction hypothesis, 
we have SLT$_{\omega}$ $\vdash$ \SEQ{\GA}{\X^i\ga}. 
Then, we obtain that
SLT$_{\omega}$ $\vdash$ \SEQ{\GA}{\X^i(\al\I\ga)}:
{\footnotesize
$$
\infer[(\I {\rm right}).]{\SEQ{\GA}{\X^i(\al\I\ga)}}{
     \infer[\mbox{\rm (we-left)}]{\SEQ{\X^i\al, \GA}{\X^i\ga}}{
          \infer*[Ind. \, hyp.]{\SEQ{\GA}{\X^i\ga}}{
          }
     }
}
$$
}
\item
Case ($\neg$I):
${\mathcal{D}}$ is of the form: 
{\footnotesize
$$
\infer[(\neg {\rm I})]{\X^i\neg\al}{
     \infer*[{\mathcal{D}}_1]{\X^j\neg\ga}{
          [\X^i\al] \GA_1
     }
     &
     \infer*[{\mathcal{D}}_2]{\X^j\ga}{
         [\X^i\al] \GA_2
     }
}
$$}where oa(${\mathcal{D}}$) = $\GA_1 \cup\GA_2$ and end(${\mathcal{D}}$) = $\X^i\neg\al$. 
By induction hypotheses, 
we have 
SLT$_{\omega}$ $\vdash$ \SEQ{\X^i\al, \GA_1}{\X^j\neg\ga} and 
SLT$_{\omega}$  $\vdash$ \SEQ{\X^i\al, \GA_2}{\X^j\ga}.
Then, we obtain that
SLT$_{\omega}$ $\vdash$ \SEQ{\GA_1, \GA_2}{\X^i\neg\al}:
{\footnotesize
$$
\infer[(\neg {\rm right}).]{\SEQ{\GA_1, \GA_2}{\X^i\neg\al}}{
   \infer[({\rm cut})]{\SEQ{\X^i\al, \GA_1, \GA_2}{}}{
        \infer*[Ind. hyp.]{\SEQ{\X^i\al, \GA_1}{\X^j\neg\ga}}{
        }
        &
        \infer[(\neg {\rm left})]{\SEQ{\X^j\neg\ga, \X^i\al, \GA_2}{}}{
             \infer*[Ind. hyp.]{\SEQ{\X^i\al, \GA_2}{\X^j\ga}}{
             }
        }
    }
}
$$
}
\item
Case (EXP): 
${\mathcal{D}}$ is of the form: 
{\footnotesize
$$
\infer[({\rm EXP})]{\be}{
     \infer*[{\mathcal{E}}_1]{\X^i\neg\al}{
          \GA_1
     }
     &
     \infer*[{\mathcal{E}}_2]{\X^i\al}{
          \GA_2
     }
}
$$}where oa(${\mathcal{D}}$) = $\GA_1 \cup\GA_2$ and end(${\mathcal{D}}$) = $\be$. 
By induction hypotheses, 
we have\linebreak SLT$_{\omega}$ $\vdash$ \SEQ{\GA_1}{\X^i\neg\al} and 
SLT$_{\omega}$ $\vdash$ \SEQ{\GA_2}{\X^i\al}.
Then, we obtain 
that 
SLT$_{\omega}$ $\vdash$ \SEQ{\GA_1, \GA_2}{\be}:
{\footnotesize
$$
\infer[(\mbox{\rm we-right}).]{\SEQ{\GA_1, \GA_2}{\be}}{
     \infer[({\rm cut})]{\SEQ{\GA_1, \GA_2}{}}{
           \infer*[Ind.hyp.]{\SEQ{\GA_2}{\X^i\al}}{
           }
           &
           \infer[({\rm cut})]{\SEQ{\X^i\al, \GA_1}{}}{
                  \infer*[Ind.\, hyp.]{\SEQ{\GA_1}{\X^i\neg\al}}{
                  }
                  &
                  \infer[(\neg {\rm left})]{\SEQ{\X^i\neg\al, \X^i\al}{}}{
                          \SEQ{\X^i\al}{\X^i\al}
                   }
            }
     }
}
$$
}
\item
Case (EXM): 
${\mathcal{D}}$ is of the form: 
{\footnotesize
$$
\infer[({\rm EXM})]{\ga}{
     \infer*[{\mathcal{E}}_1]{\ga}{
          [\X^i\neg\al] \GA_1
     }
     &
     \infer*[{\mathcal{E}}_2]{\ga}{
          [\X^i\al] \GA_2
     }
}
$$}where oa(${\mathcal{D}}$) = $\GA_1 \cup\GA_2$ and end(${\mathcal{D}}$) = $\ga$. 
By induction hypotheses, 
we have\linebreak SLT$_{\omega}$ $\vdash$ \SEQ{\X^i\neg\al, \GA_1}{\ga} and 
SLT$_{\omega}$ $\vdash$ \SEQ{\X^i\al, \GA_2}{\ga}.
Then, we obtain 
that
SLT$_{\omega}$ $\vdash$ \SEQ{\GA_1, \GA_2}{\ga}:
{\footnotesize
$$
\infer[(\mbox{\rm ex-middle}).]{\SEQ{\GA_1, \GA_2}{\ga}}{
      \infer*[(\mbox{\rm we-left})]{\SEQ{\X^i\neg\al, \GA_1, \GA_2}{\ga}}{
          \infer*[Ind.\, hyp.]{\SEQ{\X^i\neg\al, \GA_1}{\ga}}{
          }
      }
      &
      \infer*[(\mbox{\rm we-left})]{\SEQ{\X^i\al, \GA_1, \GA_2}{\ga}}{
           \infer*[Ind.\, hyp.]{\SEQ{\X^i\al, \GA_2}{\ga}}{
           }
      }
}
$$
}
\item
Case ($\G$I):
${\mathcal{D}}$ is of the form: 
{\footnotesize
$$
\infer[(\G {\rm I})]{\X^i\G\al}{
     \infer*[P_j]{\{~\X^{i+j}\al~\}_{j\in \omega}}{
          \GA_j
     }
}
$$}where oa(${\mathcal{D}}$) = $\displaystyle{\GA = \bigcup_{j\in\omega} \GA_j}$ and end(${\mathcal{D}}$) = $\X^i\G\al$. 
By induction hypotheses,
we have SLT$_{\omega}$ $\vdash$ \SEQ{\GA_j}{\X^{i+j}\al} for all $j\in\omega$. 
Then, we obtain 
that
SLT$_{\omega}$ $\vdash$ \SEQ{\GA}{\X^i\G\al}:
{\footnotesize
$$
\infer[(\G {\rm right}).]{\SEQ{\GA}{\X^i\G\al}}{
      \infer*[(\mbox{\rm we-left})]{\{~\SEQ{\GA}{\X^{i+j}\al}~\}_{j\in\omega}}{
          \infer*[Ind.hyp.]{\SEQ{\GA_j}{\X^{i+j}\al}}{
          }
      }
}
$$}Note that the induction hypothesis is applied for each of the denumerable set of premises.
\item
Case ($\F$E):  
${\mathcal{D}}$ is of the form: 
{\footnotesize
$$
\infer[(\F {\rm E})]{\ga}{
     \infer*[{\mathcal{D}}']{\X^i\F\al}{
          \GA'
     }
     &
     \infer*[{\mathcal{D}}_j]{\{~\ga~\}_{j\in\omega}}{
        [\X^{i+j}\al] \GA_j
     }
}
$$}where oa(${\mathcal{D}}$) = $\displaystyle{\GA'\cup\GA}$ with $\displaystyle{\GA = \bigcup_{j\in\omega} \GA_j}$ 
and end(${\mathcal{D}}$) = $\ga$.
By induction hypotheses,
we have 
SLT$_{\omega}$ $\vdash$ \SEQ{\GA'}{\X^i\F\al} and 
SLT$_{\omega}$ $\vdash$ \SEQ{\X^{i+j}\al, \GA_j}{\ga} for all $j\in\omega$. 
Then we obtain 
that 
SLT$_{\omega}$ $\vdash$ \SEQ{\GA', \GA}{\ga} 
by the following derivation where the induction hypothesis is applied for each of the denumerable set of premises: 
{\footnotesize
$$
\infer[({\rm cut}).]{\SEQ{\GA', \GA}{\ga}}{
\infer*[Ind. hyp.]{\SEQ{\GA'}{\X^i\F\al}}{
}
&
\infer[(\F {\rm left})]{\SEQ{\X^i\F\al, \GA}{\ga}}{
      \infer*[(\mbox{\rm we-left})]{\{~\SEQ{\X^{i+j}\al, \GA}{\ga}~\}_{j\in\omega}}{
          \infer*[Ind.hyp.]{\SEQ{\X^{i+j}\al, \GA_j}{\ga}}{
          }
      }
}
}
$$
}
\end{enumerate}

\item
We prove 2 by induction on the derivations ${\mathcal{D}}$ of \SEQ{\GA}{\be} in SLT$_{\omega}$ $-$ (cut). We distinguish the cases according to the last inference of ${\mathcal{D}}$. We show some cases. 
\begin{enumerate}

\item
Case (we-right): 
${\mathcal{D}}$ is of the form: 
{\footnotesize
$$
\infer[\mbox{\rm (we-right)}]{\SEQ{\GA}{\al}}{
  \infer*[{\mathcal{D}}']{\SEQ{\GA}{}}{
  }
}
$$}By induction hypothesis, we have a normal derivation ${\mathcal{E}}'$ in NLT$_{\omega}$ of the form: 
{\footnotesize
$$
\infer*[{\mathcal{E}}']{\neg p\LAND p}{
    \GA
}
$$}where oa(${\mathcal{E}}'$) = $\GA$ and end(${\mathcal{E}}'$) = $\neg p \LAND p$. 
Then, we obtain a required normal derivation ${\mathcal{E}}$ by: 
{\footnotesize
$$
\infer[{\rm (Exp)}]{\al}{
     \infer[{\rm (\LAND E1)}]{\neg p}{
         \infer*[{\mathcal{E}}']{\neg p\LAND p}{
               \GA
         }
     }
     &
     \infer[{\rm (\LAND E2)}]{p}{
         \infer*[{\mathcal{E}}']{\neg p\LAND p}{
               \GA
         }
     }
}
$$}where oa(${\mathcal{E}}$) = $\GA$ and end(${\mathcal{E}}$) = $\al$.

\item
Case ($\neg$left): 
${\mathcal{D}}$ is of the form: 
{\footnotesize
$$
\infer[(\neg{\rm left}).]{\SEQ{\X^i\neg\al, \GA}{}}{
  \infer*[{\mathcal{D}}']{\SEQ{\GA}{\X^i\al}}{
  }
}
$$}By induction hypothesis, we have a normal derivation ${\mathcal{E}}'$ in NLT$_{\omega}$ of the form: 
{\footnotesize
$$
\infer*[{\mathcal{E}}']{\X^i\al}{
    \GA
}
$$}where oa(${\mathcal{E}}'$) = $\GA$ and end(${\mathcal{E}}'$) = $\X^i\al$. 
Then, we obtain a required normal derivation ${\mathcal{E}}$ by: 
{\footnotesize
$$
\infer[({\rm EXP})]{\neg p\LAND p}{
       \X^i\neg\al
       &
       \infer*[{\mathcal{E}}']{\X^i\al}{
           \GA
       }
}
$$}where oa(${\mathcal{E}}$) = $\{ \X^i\neg\al \}\cup \GA$ and end(${\mathcal{E}}$) = $\neg p\LAND p$ (i.e., $\bot$).
We remark that the last inference (EXP) in ${\mathcal{E}}$ cannot be replaced with ($\I$E), because using ($\I$E) entails a possibility of developing a non-normal derivation. Namely, there is a possibility of the case that the last inference of ${\mathcal{E}}'$ is ($\I$I$^*$). 
\item
Case (ex-middle): 
${\mathcal{D}}$ is of the form: 
{\footnotesize
$$
\infer[(\mbox{\rm ex-middle}).]{\SEQ{\GA}{\ga}}{
  \infer*[{\mathcal{D}}_1]{\SEQ{\X^i\neg\al, \GA}{\ga}}{
  }
  &
  \infer*[{\mathcal{D}}_2]{\SEQ{\X^i\al, \GA}{\ga}}{
  }
}
$$}By induction hypotheses, we have normal derivations ${\mathcal{E}}_1$ and ${\mathcal{E}}_2$ in NLT$_{\omega}$ of the form: 
{\footnotesize
$$
\infer*[{\mathcal{E}}_1]{\ga}{
    \X^i\neg\al, \GA
}
\quad\quad\quad
\infer*[{\mathcal{E}}_2]{\ga}{
    \X^i\al, \GA
}
$$}where 
oa(${\mathcal{E}}_1$) = $\{ \X^i\neg\al \} \cup \GA$, 
oa(${\mathcal{E}}_2$) = $\{ \X^i\al \} \cup \GA$, 
end(${\mathcal{E}}_1$) = $\ga$, and end(${\mathcal{E}}_2$) = $\ga$.  
Then, we obtain a required normal derivation ${\mathcal{E}}$ by:  
{\footnotesize
$$
\infer[(\mbox{\rm EXM})]{\ga}{
       \infer*[{\mathcal{E}}_1]{\ga}{
            [\X^i\neg\al] \GA
       }
       &
       \infer*[{\mathcal{E}}_2]{\ga}{
            [\X^i\al] \GA
       }
}
$$}where oa(${\mathcal{E}}$) = $\GA$ and end(${\mathcal{E}}$) = $\ga$. 
\item
Case ($\F$left): 
${\mathcal{D}}$ is of the form: 
{\footnotesize
$$
\infer[(\F {\rm left}).]{\SEQ{\X^i\F\al, \GA}{\ga}}{
  \infer*[{\mathcal{D}}']{\{~\SEQ{\X^{i+k}\al, \GA}{\ga}~\}_{j\in\omega}}{
  }
}
$$}By induction hypotheses, we have normal derivations ${\mathcal{E}}_j$ for all $j\in\omega$ in NLT$_{\omega}$ of the form: 
{\footnotesize
$$
\infer*[{\mathcal{E}}_j]{\ga}{
    \X^{i+j}\al
    &
    \GA_j
}
$$}where oa(${\mathcal{E}}_j$) = $\{\X^{i+j}\al \} \cup\GA_j$ with 
$\displaystyle{\GA = \bigcup_{j\in\omega} \GA_j}$ and end(${\mathcal{E}}_j$) = $\ga$.  
Then, we obtain a required normal derivation ${\mathcal{E}}$ by:  
{\footnotesize
$$
\infer[(\F {\rm E})]{\ga}{
    \X^i\F\al
    &
    \infer*[{\mathcal{E}}_j]{\{~\ga~\}_{j\in\omega}}{
        [\X^{i+j}\al]
        &
        \GA_j
    }
}
$$}where oa(${\mathcal{E}}$) = $\{\X^i\F\al\}\cup\GA$ and end(${\mathcal{E}}$) = $\ga$. 
\end{enumerate}
\end{enumerate}
\QED


\begin{thm}[Equivalence between NLT$_{\omega}$ and SLT$_{\omega}$]
\label{equivalence-nlt-slt}
For any formula $\al$, 
{\rm SLT$_{\omega}$} $\vdash$ \SEQ{}{\al}  iff  $\al$ is 
derivable
in {\rm NLT$_{\omega}$}.
\end{thm}
\PROOF
Taking $\emptyset$ as $\GA$ in Lemma \ref{prop-nlt-slt}, we obtain the required fact. 
\QED

\begin{thm}[Normalization for NLT$_{\omega}$]
\label{normal-nlt-omega}
All derivations in {\rm NLT$_{\omega}$} are normalizable. 
More precisely, if a derivation ${\mathcal{D}}$ in {\rm NLT$_{\omega}$} is given, 
then we  obtain a normal derivation ${\mathcal{E}}$ in {\rm NLT$_{\omega}$} such that 
{\rm oa(${\mathcal{E}}$)} $=$ {\rm oa(${\mathcal{D}}$)} and 
{\rm end(${\mathcal{E}}$)} $=$ {\rm end(${\mathcal{D}}$)}. 
\end{thm}
\PROOF
Suppose that a derivation ${\mathcal{D}}$ in NLT$_{\omega}$ is given, and suppose that oa(${\mathcal{D}}$) = $\GA$ and end(${\mathcal{D}}$) = $\be$. Then, by Lemma \ref{prop-nlt-slt} (1), we obtain SLT$_{\omega}$ $\vdash$ \SEQ{\GA}{\be}. By the cut-elimination theorem for SLT$_{\omega}$, we obtain SLT$_{\omega}$ $-$ (cut) $\vdash$ \SEQ{\GA}{\be}. Then, by Lemma \ref{prop-nlt-slt} (2), we obtain a normal derivation $Q$ in NLT$_{\omega}$ such that oa(${\mathcal{E}}$) = oa(${\mathcal{D}}$) and end(${\mathcal{E}}$) = end(${\mathcal{D}}$). 
\QED

\
\section{Concluding remarks and acknowledgments}
In this paper we introduced  a unified Gentzen-style framework for the until-free propositional logic LTL.  In this framework, based on infinitary rules and rules for primitive negation,  sequent calculus and natural deduction can be treated in a uniform way, that eases a proof of their deductive equivalence and a proof of normalization for the natural deduction system. More specifically, natural deduction derivations are translated to sequent calculus derivations with cuts, and cut-free derivations are translated to normal derivations in natural deduction. In this way, cut elimination provides the bridge to an indirect proof normalization. In future work, we plan to improve the correspondence between cut elimination and normalization to a bi-directional one with the use of general elimination rules (as in \cite[Chapter~8]{NP-2001}). This should also address a question posed by one of the referees (who are gratefully acknowledged for their valuable comments) on the correspondence between steps of cut elimination and reduction steps in a normalization sequence. Other desiderata for further work  include a direct proof of normalization, and an inquiry on strong normalization and the Church-Rosser theorem.

\vskip 5pt
\noindent
This research was supported by JSPS KAKENHI Grant Number 23K10990, the project  ``Infinity and Intensionality: Towards A New Synthesis'' funded by the Research Council of Norway, and ``Modalities in Substructural Logics: Theory, Methods and Applications MOSAIC'', funded by the Community Research and Development Information Service (CORDIS) of the European Commission. The second author also acknowledges  the MIUR Excellence Department Project awarded to Dipartimento di Matematica, Universit\`a di Genova, CUP D33C23001110001 and  the ``Gruppo Nazionale per le Strutture Algebriche, Geometriche e le loro Applicazioni'' (GNSAGA) of the Istituto Nazionale di Alta Matematica (INdAM).

\nocite{*}
\bibliographystyle{eptcs}
\bibliography{generic}

\begin{thebibliography}{10}
\providecommand{\bibitemdeclare}[2]{}
\providecommand{\surnamestart}{}
\providecommand{\surnameend}{}
\providecommand{\urlprefix}{Available at }
\providecommand{\url}[1]{\texttt{#1}}
\providecommand{\href}[2]{\texttt{#2}}
\providecommand{\urlalt}[2]{\href{#1}{#2}}
\providecommand{\doi}[1]{doi:\urlalt{https://doi.org/#1}{#1}}
\providecommand{\eprint}[1]{arXiv:\urlalt{https://arxiv.org/abs/#1}{#1}}
\providecommand{\bibinfo}[2]{#2}

\bibitemdeclare{article}{AFRICK}
\bibitem{AFRICK}
\bibinfo{author}{Henry \surnamestart Africk\surnameend} (\bibinfo{year}{1992}):
  \emph{\bibinfo{title}{Classical logic, intuitionistic logic, and the Peirce
  rule}}.
\newblock {\slshape \bibinfo{journal}{Notre Dame Journal of Formal Logic 33
  (2), pp. 229-235}}, \doi{10.1305/ndjfl/1093636101}.

\bibitemdeclare{article}{AN84}
\bibitem{AN84}
\bibinfo{author}{Ahmad \surnamestart Almukdad\surnameend} \&
  \bibinfo{author}{David \surnamestart Nelson\surnameend}
  (\bibinfo{year}{1984}): \emph{\bibinfo{title}{Constructible falsity and
  inexact predicates}}.
\newblock {\slshape \bibinfo{journal}{Journal of Symbolic Logic 49 (1), pp.
  231-233}}, \doi{10.2307/2274105}.

\bibitemdeclare{article}{BM2003}
\bibitem{BM2003}
\bibinfo{author}{Stefano \surnamestart Baratella\surnameend} \&
  \bibinfo{author}{Andrea \surnamestart Masini\surnameend}
  (\bibinfo{year}{2003}): \emph{\bibinfo{title}{A proof-theoretic investigation
  of a logic of positions}}.
\newblock {\slshape \bibinfo{journal}{Annals of Pure and Applied Logic 123, pp.
  135-162}}, \doi{10.1016/S0168-0072(03)00021-6}.

\bibitemdeclare{article}{BM04}
\bibitem{BM04}
\bibinfo{author}{Stefano \surnamestart Baratella\surnameend} \&
  \bibinfo{author}{Andrea \surnamestart Masini\surnameend}
  (\bibinfo{year}{2004}): \emph{\bibinfo{title}{An approach to infinitary
  temporal proof theory}}.
\newblock {\slshape \bibinfo{journal}{Archive for Mathematical Logic 43 (8),
  pp. 965-990}}, \doi{10.1007/S00153-004-0237-Z}.

\bibitemdeclare{article}{BBGS2006}
\bibitem{BBGS2006}
\bibinfo{author}{Alexander \surnamestart Bolotov\surnameend},
  \bibinfo{author}{Artie \surnamestart Basukoski\surnameend},
  \bibinfo{author}{Oleg~M. \surnamestart Grigoriev\surnameend} \&
  \bibinfo{author}{Vasilyi \surnamestart Shangin\surnameend}
  (\bibinfo{year}{2006}): \emph{\bibinfo{title}{Natural deduction calculus for
  linear-time temporal logic}}.
\newblock {\slshape \bibinfo{journal}{Proceedings of the 10th European
  Conference on Logics in Artificial Intelligence (JELIA 2006), Lecture Notes
  in Computer Science 4160, pp. 56-68}}, \doi{10.1007/11853886\_7}.

\bibitemdeclare{article}{BS2012}
\bibitem{BS2012}
\bibinfo{author}{Alexander \surnamestart Bolotov\surnameend} \&
  \bibinfo{author}{Vasilyi \surnamestart Shangin\surnameend}
  (\bibinfo{year}{2012}): \emph{\bibinfo{title}{Natural deduction system in
  paraconsistent setting: Proof search for PCont}}.
\newblock {\slshape \bibinfo{journal}{Journal of Intelligent Systems 21 (1),
  pp. 1-24}}, \doi{10.1515/JISYS-2011-0021}.

\bibitemdeclare{article}{BN-2009}
\bibitem{BN-2009}
\bibinfo{author}{Bianca \surnamestart Boretti\surnameend} \&
  \bibinfo{author}{Sara \surnamestart Negri\surnameend} (\bibinfo{year}{2009}):
  \emph{\bibinfo{title}{Decidability for Priorean linear time using a
  fixed-point labelled calculus}}.
\newblock {\slshape \bibinfo{journal}{Proceedings of the 18th International
  Conference on Automated Reasoning with Analytic Tableaux and Related Methods
  (TABLEAUX), Lecture Notes in Computer Science 5607, pp. 108-122}},
  \doi{10.1007/978-3-642-02716-1\_9}.

\bibitemdeclare{incollection}{BN-2010}
\bibitem{BN-2010}
\bibinfo{author}{Bianca \surnamestart Boretti\surnameend} \&
  \bibinfo{author}{Sara \surnamestart Negri\surnameend} (\bibinfo{year}{2010}):
  \emph{\bibinfo{title}{On the finitization of Priorean linear time}}.
\newblock In \bibinfo{editor}{D'Agostino \surnamestart et~al.\surnameend},
  editor: {\slshape \bibinfo{booktitle}{New Essays in Logic and Philosophy of
  Science}}, \bibinfo{publisher}{College Publications},
  \bibinfo{address}{London}.

\bibitemdeclare{article}{CGP-FSCD-2023}
\bibitem{CGP-FSCD-2023}
\bibinfo{author}{Serenella \surnamestart Cerrito\surnameend},
  \bibinfo{author}{Valentin \surnamestart Goranko\surnameend} \&
  \bibinfo{author}{Sophie \surnamestart Paillocher\surnameend}
  (\bibinfo{year}{2023}): \emph{\bibinfo{title}{Partial model checking and
  partial model synthesis in LTL using a Tableau-based approach}}.
\newblock {\slshape \bibinfo{journal}{Proceedings of the 8th International
  Conference on Formal Structures for Computation and Deduction (FSCD), pp.
  23:1-23:21}}, \doi{10.4230/LIPICS.FSCD.2023.23}.

\bibitemdeclare{article}{EMERSON1990}
\bibitem{EMERSON1990}
\bibinfo{author}{E.~Allen \surnamestart Emerson\surnameend}
  (\bibinfo{year}{1990}): \emph{\bibinfo{title}{Temporal and modal logic}}.
\newblock {\slshape \bibinfo{journal}{In: Handbook of Theoretical Computer
  Science, Formal Models and Semantics (B), Jan van Leeuwen (Ed.), pp.
  995-1072, Elsevier and MIT Press}}, \doi{10.1016/B978-0-444-88074-1.50021-4}.

\bibitemdeclare{article}{GHLNO-CSL-2008}
\bibitem{GHLNO-CSL-2008}
\bibinfo{author}{Joxe \surnamestart Gaintzarain\surnameend},
  \bibinfo{author}{Montserrat \surnamestart Hermo\surnameend},
  \bibinfo{author}{Paqui \surnamestart Lucio\surnameend},
  \bibinfo{author}{Marisa \surnamestart Navarro\surnameend} \&
  \bibinfo{author}{Fernando \surnamestart Orejas\surnameend}
  (\bibinfo{year}{2007}): \emph{\bibinfo{title}{A cut-free and invariant-free
  sequent calculus for PLTL}}.
\newblock {\slshape \bibinfo{journal}{Proceedings of the 21st International
  Workshop on Computer Science Logic, Lecture Notes in Computer Science 4646,
  pp. 481-495}}, \doi{10.1007/978-3-540-74915-8\_36}.

\bibitemdeclare{article}{GENTZEN}
\bibitem{GENTZEN}
\bibinfo{author}{Gerhard \surnamestart Gentzen\surnameend}
  (\bibinfo{year}{1969}): \emph{\bibinfo{title}{Collected papers of Gerhard
  Gentzen}}.
\newblock {\slshape \bibinfo{journal}{M.E. Szabo (ed.), Studies in logic and
  the foundations of mathematics, North-Holland (English translation}},
  \doi{10.2307/2272429}.

\bibitemdeclare{article}{GUREVICH}
\bibitem{GUREVICH}
\bibinfo{author}{Yuri \surnamestart Gurevich\surnameend}
  (\bibinfo{year}{1977}): \emph{\bibinfo{title}{Intuitionistic logic with
  strong negation}}.
\newblock {\slshape \bibinfo{journal}{Studia Logica}} \bibinfo{volume}{36}, pp.
  \bibinfo{pages}{49--59}, \doi{10.1007/BF02121114}.

\bibitemdeclare{article}{BL-2008}
\bibitem{BL-2008}
\bibinfo{author}{\surnamestart {Kai Br\"unnler}\surnameend} \&
  \bibinfo{author}{Martin \surnamestart Lange\surnameend}
  (\bibinfo{year}{2008}): \emph{\bibinfo{title}{Cut-free sequent systems for
  temporal logic}}.
\newblock {\slshape \bibinfo{journal}{Journal of Logic and Algebraic Methods in
  Programming 76 (2), pp. 216-225}}, \doi{10.1016/J.JLAP.2008.02.004}.

\bibitemdeclare{article}{KAMIDEBSL06}
\bibitem{KAMIDEBSL06}
\bibinfo{author}{Norihiro \surnamestart Kamide\surnameend}
  (\bibinfo{year}{2006}): \emph{\bibinfo{title}{An equivalence between sequent
  calculi for linear-time temporal logic}}.
\newblock {\slshape \bibinfo{journal}{Bulletin of the Section of the Logic}}
  \bibinfo{volume}{35}(\bibinfo{number}{4}), pp. \bibinfo{pages}{187--194}.

\bibitemdeclare{article}{KAMIDEMSCS2015}
\bibitem{KAMIDEMSCS2015}
\bibinfo{author}{Norihiro \surnamestart Kamide\surnameend}
  (\bibinfo{year}{2015}): \emph{\bibinfo{title}{Embedding theorems for LTL and
  its variants}}.
\newblock {\slshape \bibinfo{journal}{Mathematical Structures in Computer
  Science}} \bibinfo{volume}{25}(\bibinfo{number}{1}), pp.
  \bibinfo{pages}{83--134}, \doi{10.1017/S0960129514000048}.

\bibitemdeclare{article}{KAMIDE-ISMVL-2023-ND}
\bibitem{KAMIDE-ISMVL-2023-ND}
\bibinfo{author}{Norihiro \surnamestart Kamide\surnameend}
  (\bibinfo{year}{2023}): \emph{\bibinfo{title}{Natural deduction with
  explosion and excluded middle}}.
\newblock {\slshape \bibinfo{journal}{Proceedings of the 53rd IEEE
  International Symposium on Multiple-valued Logic (ISMVL 2023)}}, pp.
  \bibinfo{pages}{24--29}, \doi{10.1109/ISMVL57333.2023.00016}.

\bibitemdeclare{misc}{KN-2023}
\bibitem{KN-2023}
\bibinfo{author}{Norihiro \surnamestart Kamide\surnameend} \&
  \bibinfo{author}{Sara \surnamestart Negri\surnameend} (\bibinfo{year}{2024}):
  \emph{\bibinfo{title}{Unified natural deduction for logics of strong
  negation}}.
\newblock \bibinfo{note}{Draft}.

\bibitemdeclare{article}{KW-FI-2011}
\bibitem{KW-FI-2011}
\bibinfo{author}{Norihiro \surnamestart Kamide\surnameend} \&
  \bibinfo{author}{Heinrich \surnamestart wansing\surnameend}
  (\bibinfo{year}{2011}): \emph{\bibinfo{title}{A paraconsistent linear-time
  temporal logic}}.
\newblock {\slshape \bibinfo{journal}{Fundamenta Informaticae 106 (1), pp.
  1-23}}, \doi{10.3233/FI-2011-374}.

\bibitemdeclare{article}{KAWAI87}
\bibitem{KAWAI87}
\bibinfo{author}{Hiroya \surnamestart Kawai\surnameend} (\bibinfo{year}{1987}):
  \emph{\bibinfo{title}{Sequential calculus for a first order infinitary
  temporal logic}}.
\newblock {\slshape \bibinfo{journal}{Zeitschrift f{\" u}r Mathematische Logik
  und Grundlagen der Mathematik 33, pp. 423-432}},
  \doi{10.1002/MALQ.19870330506}.

\bibitemdeclare{article}{NP-2001}
\bibitem{NP-2001}
\bibinfo{author}{Sara \surnamestart Negri\surnameend} \& \bibinfo{author}{Jan
  \surnamestart von Plato\surnameend} (\bibinfo{year}{2001}):
  \emph{\bibinfo{title}{Structural Proof Theory}}.
\newblock {\slshape \bibinfo{journal}{Cambridge University Press}},
  \doi{10.1017/CBO9780511527340}.

\bibitemdeclare{article}{NELSON}
\bibitem{NELSON}
\bibinfo{author}{David \surnamestart Nelson\surnameend} (\bibinfo{year}{1949}):
  \emph{\bibinfo{title}{Constructible falsity}}.
\newblock {\slshape \bibinfo{journal}{Journal of Symbolic Logic 14, pp.
  16-26}}, \doi{10.2307/2268973}.

\bibitemdeclare{article}{KP-LLP-2021}
\bibitem{KP-LLP-2021}
\bibinfo{author}{\surnamestart {Nils K\" urbis}\surnameend} \&
  \bibinfo{author}{Yaroslav \surnamestart Petrukhin\surnameend}
  (\bibinfo{year}{2021}): \emph{\bibinfo{title}{Normalisation for some quite
  interesting many-valued logics}}.
\newblock {\slshape \bibinfo{journal}{Logic and Logical Philosophy}}
  \bibinfo{volume}{30}(\bibinfo{number}{3}), pp. \bibinfo{pages}{493--534},
  \doi{10.12775/LLP.2021.009}.

\bibitemdeclare{article}{PAECH1988}
\bibitem{PAECH1988}
\bibinfo{author}{Barbara \surnamestart Paech\surnameend}
  (\bibinfo{year}{1988}): \emph{\bibinfo{title}{Gentzen-systems for
  propositional temporal logics}}.
\newblock {\slshape \bibinfo{journal}{Lecture Notes in Computer Science 385,
  pp. 240-253}}, \doi{10.1007/BFB0026305}.

\bibitemdeclare{article}{von-Plato-draft-1998}
\bibitem{von-Plato-draft-1998}
\bibinfo{author}{Jan \surnamestart von Plato\surnameend}
  (\bibinfo{year}{1999}): \emph{\bibinfo{title}{Proof theory of full classical
  propositional logic}}.
\newblock {\slshape \bibinfo{journal}{Manuscript, 16 pages}}.

\bibitemdeclare{article}{elr}
\bibitem{elr}
\bibinfo{author}{Jan \surnamestart von Plato\surnameend}
  (\bibinfo{year}{2014}): \emph{\bibinfo{title}{Elements of Logical
  Reasoning}}.
\newblock {\slshape \bibinfo{journal}{Cambridge University Press}},
  \doi{10.1017/CBO9781139567862}.

\bibitemdeclare{article}{saved}
\bibitem{saved}
\bibinfo{author}{Jan \surnamestart von Plato\surnameend}
  (\bibinfo{year}{2017}): \emph{\bibinfo{title}{Saved from the Cellar: Gerhard
  Gentzen's Shorthand Notes on Logic {\&} Foundations of Mathematics}}.
\newblock {\slshape \bibinfo{journal}{Springr}}.

\bibitemdeclare{article}{PLIUSKEVICIUS1991}
\bibitem{PLIUSKEVICIUS1991}
\bibinfo{author}{Regimantas \surnamestart {Pliu\v skevi\v cius}\surnameend}
  (\bibinfo{year}{1991}): \emph{\bibinfo{title}{Investigation of finitary
  calculus for a discrete linear time logic by means of infinitary calculus}}.
\newblock {\slshape \bibinfo{journal}{Lecture Notes in Computer Science 502,
  pp. 504-528}}, \doi{10.1007/BFB0019366}.

\bibitemdeclare{article}{PNUELI1977}
\bibitem{PNUELI1977}
\bibinfo{author}{Amir \surnamestart Pnueli\surnameend} (\bibinfo{year}{1977}):
  \emph{\bibinfo{title}{The temporal logic of programs}}.
\newblock {\slshape \bibinfo{journal}{Proceedings of the 18th IEEE Symposium on
  Foundations of Computer Science, pp. 46-57}}, \doi{10.1109/SFCS.1977.32}.

\bibitemdeclare{article}{PRAWITZ}
\bibitem{PRAWITZ}
\bibinfo{author}{Dag \surnamestart Prawitz\surnameend} (\bibinfo{year}{1965}):
  \emph{\bibinfo{title}{Natural deduction: a proof-theoretical study}}.
\newblock {\slshape \bibinfo{journal}{Almqvist and Wiksell, Stockholm}},
  \doi{10.2307/2271676}.

\bibitemdeclare{article}{Priest-2019}
\bibitem{Priest-2019}
\bibinfo{author}{G.~\surnamestart Priest\surnameend} (\bibinfo{year}{2019}):
  \emph{\bibinfo{title}{Natural deduction systems for logics in the FDE
  family}}.
\newblock {\slshape \bibinfo{journal}{New Essays on Belnap--Dunn Logic
  (Synthese Library 418), pp. 279-292}}, \doi{10.1007/978-3-030-31136-0_16}.

\bibitemdeclare{article}{SZABO1980}
\bibitem{SZABO1980}
\bibinfo{author}{Manfred~E. \surnamestart Szabo\surnameend}
  (\bibinfo{year}{1980}): \emph{\bibinfo{title}{A sequent calculus for {Kr\"
  oger} logic}}.
\newblock {\slshape \bibinfo{journal}{Lecture Notes in Computer Science 148,
  pp. 295-303}}, \doi{10.1007/3-540-11981-7\_21}.

\end{thebibliography}

\end{document}